\documentclass{iopart} 

\usepackage{latexsym,iopams} 

\eqnobysec 

\newcommand{\sgp}{{\scriptscriptstyle S}} 
\newcommand{\nep}{{\scriptscriptstyle N}} 
\newcommand{\eip}{{\scriptscriptstyle I}} 
\newcommand{\eiip}{{\scriptscriptstyle I\!I}} 
\newcommand{\KV}{\xi_{\scriptscriptstyle (p)}} 
\newcommand{\HGV}{\xi} 
\newcommand{\bg}{\bar{g}} 
\newcommand{\ba}{\bar{A}} 
\newcommand{\bder}{\bar{\nabla}} 
\newcommand{\bgauge}{\bar{\chi}} 
\newcommand{\bfs}{\bar{F}} 
\newcommand{\chg}{\mbox{\large ${\cal H}$}} 
\newcommand{\ctrext}{K} 
\newcommand{\bfld}{\bar{\psi}} 
\newcommand{\beps}{\bar{\varepsilon}_{b a c_1 \cdots c_{n-2}}} 
\newcommand{\TKV}{\xi_{\scriptscriptstyle (t)}} 
\newcommand{\AKV}{\xi_{\scriptscriptstyle (\varphi)}} 
\newcommand{\RKV}{\xi_{\scriptscriptstyle (i)}} 
\newcommand{\HKV}{\xi_{\scriptscriptstyle (h)}} 
\newcommand{\potkil}{\lambda} 
\newcommand{\odr}{\Or} 
\begin{document}

\baselineskip .15in
\begin{flushright}
WU-AP/255/06
\end{flushright}

\title{Asymptotic symmetries on Kerr--Newman horizon 
without anomaly of diffeomorphism invariance} 

\author{Jun-ichirou Koga} 

\address{Advanced Research Institute for Science and Engineering, 
Waseda University, Shinjuku, Tokyo 169-8555, Japan} 

\ead{koga@gravity.phys.waseda.ac.jp} 

%%%%%%%%%%%%%%%%%%%%%%%%%%%%%%%%%
\begin{abstract} 
We analyze asymptotic symmetries on the Killing horizon 
of the four-dimensional Kerr--Newman black hole. 
We first derive the asymptotic Killing vectors on the Killing horizon, 
which describe the asymptotic symmetries, 
and find that the general form of these asymptotic Killing vectors 
is the universal one possessed by arbitrary Killing horizons. 
We then construct the phase space associated with 
the asymptotic symmetries. 
It is shown that the phase space of an extreme black hole 
either has the size comparable with a non-extreme black hole, 
or is small enough to exclude degeneracy, 
depending on whether or not the global structure of a Killing horizon 
particular to an extreme black hole is respected. 
We also show that the central charge in the Poisson brackets algebra 
of these asymptotic symmetries vanishes, 
which implies that there is not an anomaly of diffeomorphism invariance. 
By taking into account other results in the literature, 
we argue that the vanishing central charge on a black hole horizon,  
in an effective theory, 
looks consistent with the thermal feature of a black hole. 
We furthermore argue that the vanishing central charge implies  
that there are infinitely many classical configurations 
that are associated with the same macroscopic state, 
while these configurations are distinguished physically. 
\end{abstract} 
%%%%%%%%%%%%%%%%%%%%%%%%%%%%%%%%% 

%\pacs{} 

%\maketitle 

%%%%%%%%%%%%%%%%%%%%%%%%%%%%%%%%%
\baselineskip 15pt
%%%%%%%%%%%%%%%%%%%%%%%%%%%%%%%%%
%                     Introduction 
%%%%%%%%%%%%%%%%%%%%%%%%%%%%%%%%%%%
\section{Introduction} 
\label{sec:Introduction} 

In spite of various efforts, the statistical origin of the thermal feature of 
a black hole \cite{BardeenCH73,Hawking75} still remains 
to be clarified. 
A universal framework will be necessary to understand, from a general point of view, 
what are microscopic degrees of freedom responsible for the thermal feature 
and how those degrees of freedom constitute a thermal object. 
One of possible ideas is to consider a universal and local geometric structure 
associated with existence of a black hole horizon 
and analyze whether such a geometric structure is related to 
the thermal feature of a black hole. 
Motivated by the success in the case of the B.T.Z. black hole 
\cite{Strominger98}, asymptotic symmetries on a black hole horizon have been analyzed 
in this context (see \cite{Fursaev04,Carlip06a} for recent reviews), 
but it still remains controversial whether this approach is successful, 
in particular whether 
the desirable form of 
a non-vanishing central charge, i.e., an anomaly of diffeomorphism 
invariance, arises {\it naturally} in the algebra 
associated with the asymptotic symmetries, 
as in the case of the B.T.Z. black hole. 

An idea behind the asymptotic symmetries on a black hole horizon is the possibility 
that the microscopic degrees of freedom responsible for the thermal feature 
may be described without the details of quantum theory of gravity. 
This idea does not seem absurd, because 
the microscopic degrees of freedom of \textit{standard} thermal radiation 
are indeed described within the classical electromagnetism. 
Whereas the quantization condition should be imposed in order to obtain  
the Planckian spectrum, 
we need not resort to the complete theory of quantum physics 
to study the thermodynamical state of standard thermal radiation. 
It is then expected that we may understand also the microscopic degrees of freedom 
responsible for the thermal feature of a black hole 
in the context of classical theory.  
In the case of a black hole, however, we are faced with the fact 
that the structure of a Killing horizon, which is fundamental in black hole 
thermodynamics, does not allow for a sufficient number of classical 
configurations, while a large number of microscopic states are necessary 
for a black hole to exhibit the thermal feature and possess the entropy. 
Then, a possible way is to consider 
a weakened structure of a Killing horizon, and analyze whether 
it allows for a sufficient number of classical configurations. 

In the previous work \cite{Koga-a}, a local geometric 
structure of a Killing horizon, called an asymptotic Killing horizon, 
was analyzed from the viewpoint of universality, 
by weakening the geometric structure of a Killing horizon 
without assuming 
any global structures of a spacetime or field equations. 
An asymptotic Killing horizon was defined as the pair $( H , \HGV^a )$ 
of a null hypersurface $H$ and its generator $\HGV^a$, 
i.e., the vector field that plays the same role only on $H$ as the generator of 
a Killing horizon. 
The generator of an asymptotic Killing horizon was then found to be given 
by the asymptotic Killing vectors, which describe the asymptotic symmetries 
on an asymptotic Killing horizon. 
It was thus shown that once there exists one asymptotic Killing horizon, 
there exist infinitely many asymptotic Killing horizons $( H , \HGV^a )$ 
on the same null hypersurface $H$, that is, the generator $\HGV^a$ is highly 
non-unique. 
These results show that we can indeed obtain degeneracy of classical configurations 
by weakening the structure of a Killing horizon. 
We also showed that asymptotic Killing horizons are physically distinguishable, 
not being a sort of gauge, by analyzing the behavior of the acceleration associated with 
the generators $\HGV^a$. We furthermore argued that the discrepancy between 
string theory \cite{StromingerVafa96,Horowitz96,Sen05} 
and the Euclidean approach to black hole thermodynamics 
\cite{HawkingHR95,Teitelboim95}
in the entropy of an extreme black hole will be resolved, 
if the microscopic states responsible for the thermal feature of a black hole 
are connected with the asymptotic Killing horizons. 

It then may be expected that the microscopic states 
of black hole thermodynamics are described by asymptotic Killing horizons. 
To clarify whether this is true or not, however, it is necessary 
to understand how asymptotic Killing horizons are 
described in a phase space, especially 
when we are interested in 
quantum and/or statistical physics of asymptotic Killing horizons. 
In particular, we need to 
derive the Poisson brackets algebra of the asymptotic symmetries 
in order to understand whether a non-vanishing central charge arises. 
It is important also to show that asymptotic Killing horizons can be 
regarded as degenerate from a macroscopic point of view, 
and hence as the microscopic states that 
constitute one thermodynamical state. 
An evidence for this will be provided, if we can show that 
the same values of thermodynamical variables are shared by 
all the asymptotic Killing horizons. 
In order to analyze these issues, however, 
we should specify the explicit form of the Lagrangian 
of the theory we consider. 
We will thus focus in this paper on the four-dimensional Einstein--Maxwell theory, 
and hence on the Kerr--Newman black hole spacetime, 
as an example that is basic but interesting enough. 
By considering explicitly the metric 
near the Killing horizon of the Kerr--Newman black hole, 
we can derive also the general form of the asymptotic Killing vectors 
on the Killing horizon. 
Thus, the purpose of this paper is to investigate those aspects 
associated with the asymptotic Killing vectors which are clarified by specifying 
the black hole solution and the Lagrangian, and provide a further evidence 
that the microscopic states of black hole thermodynamics 
are described by asymptotic Killing horizons. 

To analyze the asymptotic Killing vectors, 
it is reasonable to focus on regular and continuous 
asymptotic symmetries, unless convincing physical explanations 
are provided, 
which show that singularity or discontinuity of asymptotic symmetries 
is essential in black hole thermodynamics. 
Then, we will utilize in this paper 
a regular coordinate system 
to derive the asymptotic Killing vectors in a correct and well-defined manner, 
while results never depend on a coordinate system used in the analysis. 
In the next section, we will thus first review the behavior of 
the metric and the gauge potential of the electromagnetic field, 
in a regular coordinate system and a regular gauge,  
near the Killing horizon of the Kerr--Newman black hole. 
In section \ref{sec:asymvec}, we will derive 
the asymptotic Killing vectors on the Killing horizon, 
based on the behavior of the metric considered in section \ref{sec:background}.  
It will be found that the general form of the asymptotic Killing vectors 
so derived is the universal one possessed by arbitrary Killing horizons \cite{Koga-a}. 
We then construct in section \ref{sec:phasespace} 
the phase space associated with the asymptotic symmetries 
and compute the Poisson brackets between the conserved charges 
conjugate to the asymptotic Killing vectors. 
To do so, we will need to incorporate the electromagnetic 
field into the asymptotic symmetries, and it will help to observe 
an interesting feature of the  phase space of an extreme black hole. 
From the computation of the Poisson brackets, we will see that 
the central charge in the Poisson brackets algebra vanishes, and that 
this fact itself implies that all asymptotic Killing horizons 
give the same value of a thermodynamical quantity. 
Finally, we will summarize and discuss our results in section \ref{sec:discussion}. 

%%%%%%%%%%%%%%%%%%%%%%%%%%%%%%%%% 
%  Near horizon 
%%%%%%%%%%%%%%%%%%%%%%%%%%%%%%%%%
\section{Near horizon} 
\label{sec:background} 

The metric $\bg_{a b}$ and the gauge potential $\ba^{\sgp}_a$ of 
the four-dimensional Kerr-Newman black hole are written 
in the Boyer--Lindquist coordinate system $(t, r, \theta, \varphi)$ as 
\begin{eqnarray} 
d\bar{s}^2 & = & - \frac{\Delta- a^2 \sin^2 \theta}{\Sigma} \; dt^2 - 
2 \; \frac{a \sin^2 \theta (r^2 + a^2 - \Delta)}{\Sigma} \; dt \, d\varphi \nonumber \\ 
& & + \; \frac{(r^2 + a^2)^2 - \Delta \, a^2 \sin^2 \theta}{\Sigma} \sin^2 \theta \; 
d\varphi^2 + \frac{\Sigma}{\Delta} dr^2 + \Sigma \, d\theta^2 , 
\label{eqn:MetricKNBL} \\ 
\ba^{\sgp}_a & = & - \; \frac{Q \, r}{\Sigma} \, ( dt )_a + 
\frac{a \, Q \, r \sin^2 \theta}{\Sigma} \; ( d\varphi )_a , 
\label{eqn:GPotKNBL} 
\end{eqnarray} 
where the bar on a field variable indicates that it is used in the following sections 
as a background, 
and the superscript $s$ on $\ba_a$ implies that it is singular. 
The functions $\Delta$ and $\Sigma$ are defined by 
\begin{equation} 
\Delta \equiv r^2 - 2 \, M r + a^2 + Q^2 \equiv ( r - r_+ ) ( r - r_- ) , \qquad 
\Sigma \equiv r^2 + a^2 \cos^2 \theta  ,  
\end{equation} 
and $M$, $J \equiv a M$, and $Q$ are the mass, the angular momentum, 
and the electric charge of the black hole, 
while the constants $r_+$ and $r_-$ are defined as 
\begin{equation} 
r_{\pm} \equiv M \pm \sqrt{M^2 - a^2 - Q^2} \; ,  
\end{equation} 
and denote the radius of the outer and the inner horizon, respectively. 
We focus in this paper on the case of $r_+ \geq r_-$, so that 
the spacetime singularity is hidden behind an event horizon. 

In the Kerr--Newman spacetime $( M , \bg_{a b} )$, 
there exist the timelike Killing vector 
$\TKV^a \equiv \left( \partial / \partial t \right)^a$ 
and the rotational Killing vectors $\RKV^a$.  
If the black hole is not rotating ($a = 0$), $\RKV^a$ 
denotes all the Killing vectors that generate O(3) rotations, 
whereas only the axial Killing vector 
$\AKV^a \equiv \left( \partial / \partial \varphi \right)^a$ is allowed 
in the rotating case ($a \neq 0$). 
The linear combination of these Killing vectors defined by 
\begin{equation} 
\HKV^a \equiv \TKV^a + \Omega_{H} \: \AKV^a , 
\end{equation} 
where $\Omega_{H} \equiv a / ( r^2_+ + a^2 )$ is 
the angular velocity of the black hole, 
becomes null on $r = r_+$, and so it generates a Killing horizon on $r = r_+$. 
The field strength $\bfs_{a b} = \bder_a \ba_b - \bder_b \ba_a$ 
of the electromagnetic field also is invariant along  
these Killing vectors as  
\begin{equation} 
{\cal L}_{\KV} \bfs_{a b} = 0 , 
\label{eqn:FSymmetry} 
\end{equation} 
where $\KV^a$ denotes collectively the Killing vectors $\HKV^a$ and $\RKV^a$. 
Although the Lie derivatives of the gauge potential along $\KV^a$ 
need not vanish in general, the gauge potential 
$\ba^{\sgp}_a$ given by (\ref{eqn:GPotKNBL}) satisfies 
${\cal L}_{\KV} \ba^{\sgp}_a = 0$ for all the Killing vectors $\KV^a$.  

Focusing on the future horizon, 
it is convenient to consider here a double null coordinate system 
that is ``co-rotating'' with the future horizon. 
(One can of course consider the past horizon in the same manner.) 
We thus define the new coordinates $U$, $V$ and $\phi$ by\footnote{% 
The coordinate system defined by (\ref{eqn:DoubleNullCoord}) 
differs from the standard one (see, e.g., \cite{FrolovNovikov})  
in the second term on the right-hand side of the last equation, which is given here by  
$\Omega_H d V$, not $\Omega_H d t$. However, it is convenient to use 
the coordinate system adopted here, 
when we consider the metric near the future horizon 
of an extreme black hole.}  
\begin{equation} \fl 
dt = dV - \frac{r^2 + a^2}{\Delta} \; dr , \qquad 
\frac{r^2 + a^2}{\Delta} \; dr = \frac{1}{2} ( dV - dU ) , \qquad 
d \varphi = d \phi + \Omega_{H} dV - \frac{a}{\Delta} \; dr .   
\label{eqn:DoubleNullCoord} 
\end{equation} 
However, the coordinate system $( U , V , \theta , \phi )$ 
does not cover the Killing horizon. 
In the case of a non-extreme black hole $r_{+} > r_{-}$, 
we then define the Kruskal coordinates $u$ and $v$ as  
\begin{equation} 
u = - \, \frac{1}{\kappa_0} \, \exp \left[ - \kappa_0 U \right] , \qquad   
v = \frac{1}{\kappa_0} \, \exp \left[ \kappa_0 V \right] , 
\label{eqn:NullCoordNE} 
\end{equation} 
where $\kappa_0 \equiv ( r_{+} - r_{-} ) / 2 ( r^2_{+} + a^2 )$ is the surface gravity of 
the Killing horizon.  
Now, the future horizon is located at $u = 0$ 
and covered by this coordinate system smoothly. 
Indeed, near the future horizon $u = 0$, the metric $\bg_{a b}$ behaves as 
\begin{equation} 
d\bar{s}^2 = - \; 2 \; \Lambda \: du \, dv 
+ 2 \; \Xi \: v \: du \, d\phi + \Sigma_{+} \: d\theta^2 
+ \frac{R^4}{\Sigma_{+}} \sin^2 \theta \: d\phi^2 + \odr(u) , 
\label{eqn:HMetricNE} 
\end{equation} 
where the functions $\Lambda$, $\Xi$, and $\Sigma_{+}$ are defined by 
\begin{eqnarray} 
\Lambda & \equiv & \frac{1}{2} \: \exp [ - \: 2 \: \kappa_0 \: r_{+} ] \: 
\Bigl( 1 - \frac{r_{-}}{r_{+}} \Bigr)^{\alpha + 1} \frac{r^2_{+}}{r^2_{+} + a^2} \: 
( 1 - a \, \Omega_{H} \sin^2 \theta ) , 
\label{eqn:LambdaNEDef} \\ 
\Xi & \equiv & \frac{a \kappa_0}{2} \: \exp [ - \: 2 \: \kappa_0 \: r_{+} ] \: 
\Bigl( 1 - \frac{r_{-}}{r_{+}} \Bigr)^{\alpha + 1} \frac{r^2_{+}}{r^2_{+} + a^2} \; 
\sin^2 \theta ,  \qquad 
\Sigma_{+} \equiv r^2_{+} + a^2 \cos^2 \theta , 
\label{eqn:FuncNEDef} 
\end{eqnarray} 
and the constants $R$ and $\alpha$ are given by 
\begin{equation} 
R^2 \equiv r^2_{+} + a^2 , \qquad \alpha \equiv \frac{r^2_{-} + a^2}{r^2_{+} + a^2} . 
\label{eqn:R2HDef} 
\end{equation} 
As one can see from the components in the regular coordinate system 
$( u , v , \theta , \phi)$, the gauge potential $\ba^{\sgp}_a$ is singular 
on the future horizon $u = 0$. 
A regular gauge is then achieved 
by the gauge transformation 
$\ba^{\sgp}_a \rightarrow \ba^{\nep}_a = \ba^{\sgp}_a  - \bder_a \bgauge^{\nep}$ 
with 
\begin{equation} 
\bgauge^{\nep} \equiv  - \; \frac{Q \, r_{+}}{( r^2_{+} + a^2 )} \: t
= - \; \frac{Q \, r_{+}}{2 \; ( r^2_{+} + a^2 )} \: V 
- \frac{Q \, r_{+}}{2 \; ( r^2_{+} + a^2 )} \: U , 
\label{eqn:RegGaugeTrNE} 
\end{equation} 
and the new gauge potential $\ba^{\nep}_a$ is written near $u = 0$ as 
\begin{equation} 
\ba^{\nep}_a = \odr(u^0) \: ( du )_a 
+ \frac{Q \, r}{\Sigma} a \sin^2 \theta \: ( d\phi )_a + \odr(u) . 
\label{eqn:RegGaugeNE} 
\end{equation} 
For all the Killing vectors $\KV^a$, we can show 
\begin{equation} 
{\cal L}_{\KV} \ba^{\nep}_a = - \bder_a {\cal L}_{\KV} \bgauge^{\nep} = 0 , 
\label{eqn:ConstPotNE} 
\end{equation} 
and hence the gauge potential $\ba^{\nep}_a$ also obeys the isometries 
of the Kerr--Newman black hole. 
Moreover, we have 
\begin{equation} 
\HKV^c \ba^{\nep}_c = \odr(u) ,  
\label{eqn:GaugeCondNE} 
\end{equation} 
where $\HKV^a$ is written in the present coordinate system as 
\begin{equation} 
\HKV^a = \kappa_0 v \left( \frac{\partial}{\partial v} \right)^a 
- \kappa_0 u \left( \frac{\partial}{\partial u} \right)^a .   
\label{eqn:XiHexpNE} 
\end{equation} 
We can obtain (\ref{eqn:GaugeCondNE}) from (\ref{eqn:RegGaugeNE}) and (\ref{eqn:XiHexpNE}). 
However, it is important to note that (\ref{eqn:GaugeCondNE}) follows 
essentially because $\ba^{\nep}_a$ is regular and $\HKV^a$ vanishes on 
the bifurcation surface ($u = v = 0$) of the Killing horizon \cite{Wald94t}, 
and (\ref{eqn:ConstPotNE}) yields ${\cal L}_{\HKV} \HKV^c \ba^{\nep}_c = 0$. 
In other words, (\ref{eqn:GaugeCondNE}) results from the fact 
that a non-extreme black hole has a bifurcate Killing horizon. 

When the black hole is extreme $r_{+} = r_{-}$, 
we introduce null coordinates $u$ and $v$ by  
\begin{equation} 
U = - \; 2 \; r_{+} \; f( - \, r_{+}^{- 1} u ) , \qquad   
V = v , 
\label{eqn:NullCoordEX} 
\end{equation} 
where the function $f(x)$ is defined as 
\begin{equation} 
f(x) \equiv x + 2 \ln x - \frac{\gamma^2}{x} , 
\end{equation} 
and the constant $\gamma$ is given by 
\begin{equation} 
\gamma^2 \equiv \frac{R^2}{r^2_{+}} = 1 + \frac{a^2}{r^2_{+}} .  
\end{equation} 
The future horizon is found to be located at $u = 0$ in this coordinate system, 
and the metric $\bg_{a b}$ behaves smoothly near $u = 0$ as 
\begin{equation} 
d\bar{s}^2 = - 2 \; \Lambda \; du \: dv + 2 \; \Xi \; du \: d\phi 
+ \Sigma_{+} \; d\theta^2 
+ \frac{R^4}{\Sigma_{+}} \sin^2 \theta \; d\phi^2 + \odr(u) , 
\label{eqn:HMetricEX}
\end{equation} 
where $\Lambda$ and $\Xi$ are defined in the extreme case as 
\begin{equation} 
\Lambda \equiv 1 - a \, \Omega_{H} \sin^2 \theta , \qquad 
\Xi \equiv a \sin^2 \theta , 
\label{eqn:FuncExDef}
\end{equation} 
while $\Sigma_{+}$ and $R$ are defined by 
(\ref{eqn:FuncNEDef}) and (\ref{eqn:R2HDef}), respectively, 
as in the non-extreme case. 
Also in the case of an extreme black hole, 
the gauge potential $\ba^{\sgp}_a$ is singular on $u = 0$, and  
we obtain a regular gauge potential $\ba^{\eip}_a$ by performing 
the gauge transformation 
$\ba^{\sgp}_a \rightarrow \ba^{\eip}_a = \ba^{\sgp}_a - \bder_a \bgauge^{\eip}$,  
where $\bgauge^{\eip}$ is given as 
\begin{equation} \fl 
\bgauge^{\eip} \equiv - \; \frac{Q \, r_{+}}{2 \: ( r^2_{+} + a^2 )} \: U 
- \frac{Q \, r_{+}}{2 \: ( r^2_{+} + a^2 )} \: V 
- Q \, r^2_{+} \frac{r^2_{+} - a^2}{( r^2_{+} + a^2 )^2} \; \rho( - \, r_{+}^{- 1} u ) 
+ \frac{Q}{2} \, \frac{r^2_+ - a^2}{( r^2_+ + a^2)^2} \: u \: v , 
\label{eqn:RegGaugeExI}
\end{equation} 
and $\rho(x)$ is defined by 
\begin{equation} 
\rho(x) \equiv \frac{1}{2} x^2 + 2 x + \gamma^2 \ln x \; . 
\end{equation} 
(One might expect that a regular gauge is achieved by 
the same gauge transformation (\ref{eqn:RegGaugeTrNE}) 
as in the non-extreme case, since the gauge potential takes the same form when 
expressed in the coordinate system $(U, V, \theta, \phi)$. 
However, it cannot remove all the singular parts of $\ba^{\sgp}_a$ in the case 
of an extreme black hole.)
The new gauge potential $\ba^{\eip}_a$ behaves near $u = 0$ as 
\begin{equation} 
\ba^{\eip}_a = \odr( u^0 ) \: ( du )_a  
+ \frac{Q \, r_{+}}{\Sigma_{+}} \, a \sin^2 \theta \: ( d\phi )_a + \odr(u) . 
\end{equation} 
However, the gauge transformation 
$\ba^{\sgp}_a \rightarrow \ba^{\eiip}_a 
= \ba^{\sgp}_a - \bder_a \bgauge^{\eiip}$, where  
\begin{equation} 
\bgauge^{\eiip} \equiv - \frac{Q \, r_{+}}{2 \: ( r^2_{+} + a^2 )} \: U 
- Q \, r^2_{+} \frac{r^2_{+} - a^2}{( r^2_{+} + a^2 )^2} \; \rho( - \, r_{+}^{- 1} u ) 
+ \frac{Q}{2} \frac{r^2_+ - a^2}{( r^2_+ + a^2 )^2} \: u \: v , 
\label{eqn:RegGaugeExII} 
\end{equation} 
also yields a regular gauge potential $\ba^{\eiip}_a$, which is written near $u = 0$ as 
\begin{equation} 
\ba^{\eiip}_a = - \frac{Q \, r_{+}}{2 \: ( r^2_{+} + a^2 )} \: ( dv )_a 
+ \odr( u^0 ) \: ( du )_a + \frac{Q \, r_+}{\Sigma_+} \, a \sin^2 \theta \: ( d \phi )_a 
+ \odr(u) . 
\end{equation} 
Therefore, both of the gauge potentials, $\ba^{\eip}_a$ and $\ba^{\eiip}_a$, 
are regular all over the future horizon. 
We also note that we have 
\begin{equation} 
{\cal L}_{\HKV} \ba_a = - \bder_a {\cal L}_{\HKV} \bgauge  = \odr(u) ,  
\label{eqn:SymGaugeEx} 
\end{equation} 
where $\ba_a$ and $\bgauge$ denote $\ba^{\eip}_a$ and $\bgauge^{\eip}$, 
or $\ba^{\eiip}_a$ and $\bgauge^{\eiip}$, respectively, 
and $\HKV^a$ is expressed in the present coordinate system as 
\begin{equation} 
\HKV^a = \left( \frac{\partial}{\partial v} \right)^a 
+ \frac{1}{2} \frac{u^2}{( u - r_{+} )^2 + a^2} 
\left( \frac{\partial}{\partial u} \right)^a .  
\label{eqn:XiHexpEX}
\end{equation} 
Since the Lie derivatives of $\ba_a^{\eip}$ and $\ba_a^{\eiip}$ 
along the rotational Killing vectors $\RKV^a$ vanish, 
all the isometries in the Kerr--Newman spacetime are respected 
by both of $\ba^{\eip}_a$ and $\ba^{\eiip}_a$ at least on the future horizon $u = 0$ as 
\begin{equation} 
{\cal L}_{\KV} \ba_a^{\eip} = \odr(u) , \qquad {\cal L}_{\KV} \ba_a^{\eiip} = \odr(u) . 
\label{eqn:ConstPotEx} 
\end{equation} 
On the other hand, 
the difference between these two gauge potentials, which is important below, 
is described as  
\begin{equation} 
\HKV^c \ba^{\eip}_c = \odr(u)  , \qquad 
\HKV^c \ba^{\eiip}_c = - \frac{Q \, r_{+}}{2 \: ( r^2_{+} + a^2 )} + \odr(u) . 
\label{eqn:ExGaugeBC} 
\end{equation}
We note that $\HKV^c \ba^{\eip}_c$ vanishes on the future horizon $u = 0$, 
whereas $\HKV^c \ba^{\eiip}_c$ does not. Thus, 
$\ba^{\eip}_a$ possesses the same property as $\ba^{\nep}_a$ 
in the non-extreme case, which satisfies (\ref{eqn:GaugeCondNE}), 
while $\ba^{\eiip}_a$ does not possess this property. 
Although $\HKV^c \ba^{\nep}_c$ in the non-extreme case is required to vanish 
on the horizon in a regular gauge, the corresponding quantity need not vanish 
in the case of an extreme black hole. This is because the Killing horizon of 
an extreme black hole does not 
have a bifurcation surface. 
Then, there are no reasons to prefer to the gauge $\ba^{\eip}_a$ and exclude 
the gauge $\ba^{\eiip}_a$. 
Rather, it is $\ba^{\eiip}_a$ that reflects faithfully the global geometric structure 
of a Killing horizon particular to an extreme black hole, while $\ba^{\eip}_a$ 
imitates the gauge in the non-extreme case. 

%%%%%%%%%%%%%%%%%%%%%%%%%%%%%%%%% 
%  Asymptotic Killing vectors 
%%%%%%%%%%%%%%%%%%%%%%%%%%%%%%%%%
\section{Asymptotic Killing vectors} 
\label{sec:asymvec} 

In the previous work \cite{Koga-a}, an asymptotic Killing vector on 
an asymptotic Killing horizon was defined based on the feature 
of the standard asymptotic symmetries, such as the B.M.S. group \cite{BMS} 
and the conformal group in a 3-dimensional asymptotically anti-de Sitter 
spacetime \cite{BrownHenneaux}, where the leading behavior of 
the asymptotic metrics is left unchanged under the asymptotic symmetry 
transformations, no matter what their sub-leading behavior is. 
In particular, an 
asymptotic Killing vector on a Killing horizon $H$ 
in a spacetime $( M , \bg_{a b} )$ is defined to be a smooth vector 
$\zeta^a$ that satisfies 
\begin{equation} 
{\cal L}_{\zeta} g_{a b} = \odr( \sigma ) , 
\label{eqn:AsymKilEq} 
\end{equation} 
for any smooth scalar $\sigma$ such that $\sigma = 0$ and $\bder_a \sigma \neq 0$ 
on $H$, where $g_{a b}$ are arbitrary smooth metrics 
that coincide with $\bg_{a b}$ on $H$ as 
\begin{equation} 
g_{a b} = \bg_{a b} + \odr( \sigma ) .  
\label{eqn:PerturbMetric} 
\end{equation} 
Thus, an asymptotic Killing vector behaves as if a Killing vector only on $H$, 
and hence it is a generalized notion of a Killing vector. 
We note also that the infinitesimal diffeomorphisms 
$g_{a b} \rightarrow g_{a b} + \mathcal{L}_{\zeta} g_{a b}$ 
along asymptotic Killing vectors $\zeta^a$ leave the form (\ref{eqn:PerturbMetric}) 
of $g_{a b}$ unchanged, which therefore generate the asymptotic symmetry group. 

We now derive explicitly the general form of the asymptotic Killing vectors $\zeta^a$ 
on the future Killing horizon of the Kerr--Newman black hole, 
by solving (\ref{eqn:AsymKilEq}).  
To do so, here we specifically take, 
as the scalar function $\sigma$ in (\ref{eqn:AsymKilEq}), 
the retarded null coordinates $u$ introduced by (\ref{eqn:NullCoordNE}) 
in the non-extreme case and (\ref{eqn:NullCoordEX}) in the extreme case, 
because these functions are smooth, vanishing, and non-degenerate 
on the future horizon.    
We then substitute the asymptotic form of the metric, 
(\ref{eqn:HMetricNE}) in the non-extreme case 
and (\ref{eqn:HMetricEX}) in the extreme case, 
for $\bg_{a b}$ in (\ref{eqn:PerturbMetric}), 
and expand the asymptotic Killing vector $\zeta^a$ as 
\begin{equation}
\zeta^a = \zeta^a_{\scriptscriptstyle (0)}(v, \theta, \phi) + 
u \; \zeta^a_{\scriptscriptstyle (1)}(v, \theta, \phi) 
+ \odr(u^{2}) .   
\label{eqn:ZetaExpand} 
\end{equation} 
The requirement that (\ref{eqn:AsymKilEq}) holds for arbitrary perturbed metrics 
$g_{a b}$ of the form (\ref{eqn:PerturbMetric}) gives ten equations for 
the components of $\zeta^a$, and they are classified into two sets of equations, 
which decouple from each other, and one equation given as   
\begin{equation} 
\zeta^{u}_{\scriptscriptstyle (0)} = 0 .  
\label{eqn:AsKilCondNEvv}
\end{equation} 

Among these two sets of equations, 
the first set consists of equations for the leading order of 
the angular components $\zeta^{\theta}_{\scriptscriptstyle (0)}$ and 
$\zeta^{\phi}_{\scriptscriptstyle (0)}$, 
whose explicit forms are given by 
\begin{eqnarray} 
& & \partial_{v} \zeta^{\theta}_{\scriptscriptstyle (0)} = 0 , 
\label{eqn:AsKilCondNEvtAp} \\ 
& & \partial_{v} \zeta^{\phi}_{\scriptscriptstyle (0)} = 0 , 
\label{eqn:AsKilCondNEvpAp} \\ 
& & \zeta^{\theta}_{\scriptscriptstyle (0)} \, \partial_{\theta} \Sigma_+ 
+ 2 \Sigma_+ \, \partial_{\theta} \zeta^{\theta}_{\scriptscriptstyle (0)} 
= 0 , 
\label{eqn:AsKilCondNEttAp} \\ 
& & \Sigma_+ \, \partial_{\phi} \zeta^{\theta}_{\scriptscriptstyle (0)} 
+ \frac{R^4}{\Sigma_+} \sin^2 \theta \: 
\partial_{\theta} \zeta^{\phi}_{\scriptscriptstyle (0)} = 0 , 
\label{eqn:AsKilCondNEtpAp} \\ 
& & \zeta^{\theta}_{\scriptscriptstyle (0)} \, \partial_{\theta} \Bigr( 
\frac{R^4}{\Sigma_+} \sin^2 \theta \Bigr) 
+ 2 \frac{R^4}{\Sigma_+} \sin^2 \theta \: 
\partial_{\phi} \zeta^{\phi}_{\scriptscriptstyle (0)} = 0 ,  
\label{eqn:AsKilCondNEppAp} 
\end{eqnarray} 
in both the non-extreme and the extreme cases. 
We note from (\ref{eqn:AsKilCondNEvtAp}) and 
(\ref{eqn:AsKilCondNEvpAp}) that $\zeta^{\theta}_{\scriptscriptstyle (0)}$ and 
$\zeta^{\phi}_{\scriptscriptstyle (0)}$ do not depend on $v$, and hence 
(\ref{eqn:AsKilCondNEttAp})--(\ref{eqn:AsKilCondNEppAp}) 
are purely two-dimensional equations. 
Indeed, they are nothing but the components of the two-dimensional 
Killing equation on a horizon sphere (a cross-section of the future Killing horizon) 
with $v$ fixed,  
and therefore the exact rotational Killing vectors $\RKV^a$ 
satisfy these equations. 
To see whether or not other solutions (``enhanced symmetries'' only at the horizon) 
happen to exist, 
we employ the axial symmetry of the Kerr--Newman spacetime and write  
$\zeta^{\theta}_{\scriptscriptstyle (0)}$ and $\zeta^{\phi}_{\scriptscriptstyle (0)}$ 
as 
\begin{equation} 
\zeta^{\theta}_{\scriptscriptstyle (0)} = P(\theta) \: e^{i m \phi} , \qquad  
\zeta^{\phi}_{\scriptscriptstyle (0)} = Q(\theta) \: e^{i m \phi} . 
\label{eqn:ZetaAngExpand} 
\end{equation} 
By substituting (\ref{eqn:ZetaAngExpand}), we rewrite 
(\ref{eqn:AsKilCondNEttAp})--(\ref{eqn:AsKilCondNEppAp}) as 
\begin{eqnarray} 
& & P(\theta) \: \partial_{\theta} \Sigma_+ 
+ 2 \Sigma_+ \, \partial_{\theta} P(\theta) = 0 , 
\label{eqn:RotComptt} \\ 
& & i \, m \, \Sigma^2_+ \, P(\theta) 
+ R^4 \sin^2 \theta \: \partial_{\theta} Q(\theta) = 0 , 
\label{eqn:RotComptp} \\ 
& & P(\theta) \, \frac{\Sigma_+}{\sin^2 \theta} \: \partial_{\theta} 
\Bigr( \frac{1}{\Sigma_+} \sin^2 \theta \Bigr) 
= - 2 \, i \, m \, Q(\theta)  , 
\label{eqn:RotComppp}
\end{eqnarray} 
respectively. 
The solution for $m = 0$ is then given by 
$P(\theta) = 0$ and $Q(\theta) = const.$, and hence it is expressed, 
by using $\partial / \partial \varphi = \partial / \partial \phi$, as 
a constant multiple of the axial Killing vector $\AKV^a$. 
For $m \neq 0$, on the other hand, we integrate 
(\ref{eqn:RotComptt}) and substitute it into  
(\ref{eqn:RotComppp}), which gives $P(\theta)$ and $Q(\theta)$. 
When they are substituted further into (\ref{eqn:RotComptp}), we obtain 
\begin{equation} 
m^2 \Sigma^{4}_+(\theta) 
+ ( {r_{+}}^2 + a^2 )^3 \Bigl( 3 a^2 \sin^2 \theta \cos^2 \theta 
- \Sigma_+(\theta) \Bigr) = 0 ,  
\label{eqn:PQCondition} 
\end{equation} 
unless $P(\theta)$ and $Q(\theta)$ are trivial as $P(\theta) = Q(\theta) = 0$, i.e., 
$\zeta^{\theta}_{\scriptscriptstyle (0)} = \zeta^{\phi}_{\scriptscriptstyle (0)} = 0$.  
It is then straightforward to show that non-trivial solutions 
of $P(\theta)$ and $Q(\theta)$  
for $m \neq 0$ do not exist when $a \neq 0$. 
Actually, (\ref{eqn:PQCondition}) gives $m^2 = 1$ when evaluated at 
$\theta = 0$, and it gives $a^2 = 0$ 
when it is evaluated at $\theta = \pi / 2$ and $m^2 = 1$ is substituted.
Therefore, a non-trivial solution in the case of $a \neq 0$ is possible only for $m = 0$, 
and is given by a constant multiple of $\AKV^a$. 
On the other hand, when $a = 0$, the spacetime is spherically symmetric, 
and hence the three exact Killing vectors of O(3) rotations are allowed.    
Moreover, in this case, any solutions should be expressed as linear combinations of 
these three Killing vectors with constant coefficients, 
because (\ref{eqn:AsKilCondNEttAp})--(\ref{eqn:AsKilCondNEppAp}) are 
the components of the two-dimensional Killing equation. 
Thus, in any case, we see that the solutions of 
$\zeta^{\theta}_{\scriptscriptstyle (0)}$ and $\zeta^{\phi}_{\scriptscriptstyle (0)}$ 
are described by  
\begin{equation} 
\zeta^{\theta}_{\scriptscriptstyle (0)} \left( \frac{\partial}{\partial \theta} \right)^a
+ \zeta^{\phi}_{\scriptscriptstyle (0)} \left( \frac{\partial}{\partial \phi} \right)^a 
= a^{\scriptscriptstyle (i)} \, \RKV^a  , 
\label{eqn:ZetaAngSol} 
\end{equation}  
where the coefficients $a^{\scriptscriptstyle (i)}$ are arbitrary constants 
and summation over the index $(i)$ is understood, if necessary. 

On the other hand, the second set of equations is written in the matrix form as 
\begin{eqnarray} 
& & \left( \begin{array}{cccc} 
0 & - \Lambda & 0 & v \: \Xi \\
- \Lambda & 0 & 0 & 0 \\ 
0 & 0 & \Sigma_+ & 0 \\
v \: \Xi & 0 & 0 & 
R^4 \sin^2 \theta / \Sigma_+ 
\end{array} \right) \left( \begin{array}{c} 
\zeta^{u}_{\scriptscriptstyle (1)} \\ 
\zeta^{v}_{\scriptscriptstyle (1)}  \\ 
\zeta^{\theta}_{\scriptscriptstyle (1)} \\ 
\zeta^{\phi}_{\scriptscriptstyle (1)} 
\end{array} \right) 
\nonumber \\ 
& = & 
\left( \begin{array}{c} 
0 \\ 
\kappa_0 \, \hat{\omega} \, \Lambda \\ 
0 \\ 
- \kappa_0 \, v \, \hat{\omega} \, \Xi 
\end{array} \right) 
+ \left( \begin{array}{c} 
0 \\ 
\kappa_0 v \, \Lambda \: \partial_{v} \hat{\omega} \\ 
\kappa_0 v \, \Lambda \: \partial_{\theta} \hat{\omega} \\ 
\kappa_0 v \, \Lambda \: \partial_{\phi} \hat{\omega} 
\end{array} \right) 
+ \left( \begin{array}{c} 
0 \\ 
\zeta^{\theta}_{\scriptscriptstyle (0)} \, \partial_{\theta} \Lambda \\ 
- v \, \Xi \, \partial_{\theta} \zeta^{\phi}_{\scriptscriptstyle (0)} \\ 
- v \, \zeta^{\theta}_{\scriptscriptstyle (0)} \, \partial_{\theta} \Xi 
- v \, \Xi \, \partial_{\phi} \zeta^{\phi}_{\scriptscriptstyle (0)} 
\end{array} \right) , 
\label{eqn:AsKilMatrixNEPre} 
\end{eqnarray} 
in the non-extreme case, and as 
\begin{equation} 
\fl 
\left( \begin{array}{cccc} 
0 & - \Lambda & 0 & \Xi \\
- \Lambda & 0 & 0 & 0 \\ 
0 & 0 & \Sigma_+ & 0 \\
\Xi & 0 & 0 & R^4 \sin^2 \theta / \Sigma_+ 
 \end{array} \right) 
\left( \begin{array}{c} 
\zeta^{u}_{\scriptscriptstyle (1)} \\ 
\zeta^{v}_{\scriptscriptstyle (1)}  \\ 
\zeta^{\theta}_{\scriptscriptstyle (1)} \\ 
\zeta^{\phi}_{\scriptscriptstyle (1)} 
\end{array} 
\right) 
= \left( \begin{array}{c} 
0 \\ 
\Lambda \: \partial_{v} \hat{\omega} \\ 
\Lambda \: \partial_{\theta} \hat{\omega} \\ 
\Lambda \: \partial_{\phi} \hat{\omega} 
\end{array} \right) 
+ \left( \begin{array}{c} 
0 \\ 
\zeta^{\theta}_{\scriptscriptstyle (0)} \, \partial_{\theta} \Lambda \\ 
- \Xi \, \partial_{\theta} \zeta^{\phi}_{\scriptscriptstyle (0)} \\ 
- \zeta^{\theta}_{\scriptscriptstyle (0)} \, \partial_{\theta} \Xi 
- \Xi \, \partial_{\phi} \zeta^{\phi}_{\scriptscriptstyle (0)} 
\end{array} \right) , 
\label{eqn:AsKilMatrixEXPre} 
\end{equation} 
in the extreme case. The function $\hat{\omega}$ in (\ref{eqn:AsKilMatrixNEPre}) 
and (\ref{eqn:AsKilMatrixEXPre}) is defined by  
\begin{equation} 
\hat{\omega}(v, \theta, \phi) \equiv 
\frac{1}{\kappa_0 v} \, \zeta^{v}_{\scriptscriptstyle (0)}(v, \theta, \phi) , 
\label{eqn:OmegaDefNE} 
\end{equation} 
in the non-extreme case, and by 
\begin{equation} 
\hat{\omega}(v, \theta, \phi) \equiv \zeta^{v}_{\scriptscriptstyle (0)}(v, \theta, \phi) , 
\label{eqn:OmegaDefEX}
\end{equation}  
in the extreme case, while it is left undetermined and hence arbitrary in what follows. 
We note that it is possible to write 
(\ref{eqn:OmegaDefNE}) and (\ref{eqn:OmegaDefEX}) in a unified manner as 
\begin{equation} 
\zeta^{v}_{\scriptscriptstyle (0)} 
= \hat{\omega} \, \HKV^v , 
\label{eqn:ZetaVKill} 
\end{equation} 
by using (\ref{eqn:XiHexpNE}) and (\ref{eqn:XiHexpEX}).  
We also note from (\ref{eqn:ZetaAngSol}) 
that we have $\zeta^{\theta}_{\scriptscriptstyle (0)} = 0$ and 
$\zeta^{\phi}_{\scriptscriptstyle (0)} = const.$ in the case of $a \neq 0$. 
On the other hand, when $a = 0$, we see from (\ref{eqn:LambdaNEDef}), 
(\ref{eqn:FuncNEDef}) and (\ref{eqn:FuncExDef})  
that $\Lambda = const.$ and $\Xi = 0$. 
Thus, in any case, the last terms 
on the right-hand side of (\ref{eqn:AsKilMatrixNEPre}) and 
(\ref{eqn:AsKilMatrixEXPre}) vanish. 
In addition, the components of the matrices on the left-hand side of 
(\ref{eqn:AsKilMatrixNEPre}) and (\ref{eqn:AsKilMatrixEXPre}) are 
the leading order of the components of $\bg_{a b}$,  
and the components of the first term on the right-hand side of 
(\ref{eqn:AsKilMatrixNEPre}) are given by the leading order of 
$- \kappa_0 \, \hat{\omega} \, \bg_{a b} \, \delta^b_u$. 
Then, by multiplying (\ref{eqn:AsKilMatrixNEPre}) and (\ref{eqn:AsKilMatrixEXPre}) 
by $u \, \bg^{c a}$, and using the explicit form (\ref{eqn:XiHexpNE}) 
of the Killing vector $\HKV^a$ in the non-extreme case, 
we obtain 
\begin{equation} 
u \, \zeta^{c}_{\scriptscriptstyle (1)} = \hat{\omega} \, \delta^{c}_{u} \, 
\HKV^u + u \, \kappa_0 \, v \, \Lambda \, \bder^{c} \hat{\omega} + \odr(u^2) , 
\label{eqn:SecondSolNE} 
\end{equation} 
in the non-extreme case, and 
\begin{equation} 
u \, \zeta^{c}_{\scriptscriptstyle (1)} = u \, \Lambda \, \bder^{c} \hat{\omega} 
+ \odr(u^2) , 
\label{eqn:SecondSolEX} 
\end{equation} 
in the extreme case. 
Therefore, by collecting (\ref{eqn:AsKilCondNEvv}),  (\ref{eqn:ZetaVKill}), 
(\ref{eqn:SecondSolNE}), and (\ref{eqn:SecondSolEX}) all together, 
adding the independent solution (\ref{eqn:ZetaAngSol}), 
and noting from (\ref{eqn:XiHexpEX}) that $\HKV^u$ in the extreme case is $\odr(u^2)$, 
we see that the general solution of $\zeta^a$ is written as  
\begin{equation}
\zeta^a = \hat{\omega} \; \HKV^a 
+ a^{\scriptscriptstyle (i)} \, \RKV^a 
+ u \, X^a + \odr(u^2) , 
\label{eqn:ZetaCov} 
\end{equation} 
where $X^a$ is given by 
\begin{equation} 
X^a \equiv \kappa_0 \, v \, \Lambda  \bder^a \hat{\omega} ,  
\label{eqn:XNEDef}
\end{equation} 
in the non-extreme case, and by 
\begin{equation} 
X^a \equiv \Lambda \bder^a \hat{\omega} .  
\label{eqn:XEXDef}
\end{equation} 
in the extreme case. 

We find that (\ref{eqn:ZetaCov}) coincides with  
the result in \cite{Koga01} in the spherically symmetric case.  
It is also possible to rewrite (\ref{eqn:ZetaCov}) into 
the form presented in \cite{Koga-a}, 
which manifestly shows that (\ref{eqn:ZetaCov}) is defined covariantly. 
To see this, we recall that the normal $\bder_a u$ to the future horizon $u = 0$ 
does not vanish, and that the future horizon is a null hypersurface. Therefore,  
$\bder^a u$ is proportional to $\HKV^a$ on $u = 0$, and hence 
there exists a smooth scalar $n$ such that 
\begin{equation} 
\HKV^a = n \bder^a u + \odr(u) = \bder^a ( n u ) + \odr(u) .  
\label{eqn:PropXiDerU} 
\end{equation} 
Actually, $n$ is given by  
$n = - \kappa_0 \, v \, \Lambda + \odr(u)$ 
in the non-extreme case, and 
$n = - \Lambda + \odr(u)$ 
in the extreme case. 
Then, by defining a scalar $\potkil$ as $\potkil = n u$, 
which is found from (\ref{eqn:PropXiDerU}) to satisfy 
$\HKV^a = \bder^a \potkil + \odr(u)$ and hence is called the potential of 
$\HKV^a$ \cite{Koga-a}, we see from (\ref{eqn:XNEDef}) and 
(\ref{eqn:XEXDef}) that we can write as 
\begin{equation} 
u X^a = - u n \bder^a \hat{\omega} + \odr(u^2) 
= - \potkil \bder^a \hat{\omega} + \odr(u^2) , 
\label{eqn:CovDefOmega} 
\end{equation} 
in both the non-extreme and the extreme cases. 
Furthermore, although $\hat{\omega}$ 
has been introduced in (\ref{eqn:OmegaDefNE}) and (\ref{eqn:OmegaDefEX}) 
as a function of $v$, $\theta$, and $\phi$ only, 
it can indeed depend on $u$, as well. 
By expanding an arbitrary smooth function $\omega(u, v, \theta, \phi)$, 
which depends also on $u$, into the Taylor series near $u = 0$ as 
\begin{equation} 
\omega(u, v, \theta, \phi) = \hat{\omega}(v, \theta, \phi) + u \, \omega'(v, \theta, \phi) 
+ \odr(u^2) , 
\end{equation} 
and using (\ref{eqn:PropXiDerU}), we obtain   
\begin{equation} 
\omega \: \HKV^a - u n \bder^a \omega 
= \hat{\omega} \: \HKV^a - u n \bder^a \hat{\omega} + \odr(u^2) .    
\label{eqn:UDependenceOm} 
\end{equation} 
Thus, dependence of $\omega$ on $u$ cancels out automatically to linear order in $u$,  
and then $\zeta^a$ is eventually written as 
\begin{equation} 
\zeta^a = \omega \, \HKV^a + a^{\scriptscriptstyle (i)} \RKV^a 
- \potkil \bder^a \omega + \odr(u^2) . 
\label{eqn:ZetaGenDefPre} 
\end{equation} 
This is indeed the form of the asymptotic Killing vectors 
presented in \cite{Koga-a}, 
which was shown to satisfy (\ref{eqn:AsymKilEq}) 
independently of the choice of $\sigma$. 
Therefore, the general form, i.e., the only possible form, of the asymptotic Killing vectors on the Killing horizon of the Kerr--Newman black hole is found to be given 
by (\ref{eqn:ZetaGenDefPre}). 
An important feature of the asymptotic Killing vectors (\ref{eqn:ZetaGenDefPre}) 
is that they exist universally on arbitrary Killing horizons \cite{Koga-a}, not only on 
the Killing horizon of the Kerr--Newman black hole.

%%%%%%%%%%%%%%%%%%%%%%%%%%%%%%%%% 
%  Covariant phase space
%%%%%%%%%%%%%%%%%%%%%%%%%%%%%%%%%
\section{Covariant phase space} 
\label{sec:phasespace} 

%%%%%%%%%%%%%%%%%%%%%%%%%%%%%%%%%%%%%%%%
\subsection{Symplectic formalism} 

Although the asymptotic Killing vectors are defined to act 
on the metric $g_{a b}$ in a spacetime, it is important to consider 
also how the actions of the asymptotic Killing vectors are represented 
in a phase space. To analyze this issue, we employ 
the symplectic (covariant phase space) formalism 
\cite{CovariantPhase,FirstLaw,WaldZoupas}, 
which we here briefly review. 

We write as $L(\psi)$ the Lagrangian of an $n$-dimensional theory, 
where $\psi^I$ denotes the dynamical field variables collectively. 
The actions ${\cal L}_{\zeta} \psi^I$ of diffeomorphisms along arbitrary vectors 
$\zeta^a$ are then identified with the variations $\delta_{\zeta} \psi^I$ 
induced by ${\cal L}_{\zeta} \psi^I$ in the covariant phase space, 
i.e., $\delta_{\zeta} \psi^I = {\cal L}_{\zeta} \psi^I$. 
In the covariant phase space, on the other hand, the variations $\delta_{\zeta} \psi^I$ 
are generated through the Poisson brackets between $\psi^I$ 
and the conserved charges $\chg[\psi, \zeta]$ 
conjugate to the vectors $\zeta^a$, 
which are defined by their derivatives $\delta \chg[\psi, \zeta]$ as 
\begin{equation} 
\delta \chg[\psi, \zeta] \equiv 
\int_{\cal C} \omega_{c_1 \cdots c_{n-1}}
(\psi; \delta_{\zeta} \psi, \delta \psi) .   
\label{eqn:ChargeDef} 
\end{equation} 
Here, ${\cal C}$ stands for an arbitrary (partial) Cauchy surface, 
on which the conserved charges $\chg[\psi, \zeta]$ are defined, 
and $\omega_{c_1 \cdots c_{n-1}}(\psi; \delta_1 \psi, \delta_2 \psi)$ 
is the symplectic current density constructed as 
\begin{equation} 
\omega_{c_1 \cdots c_{n-1}}(\psi; \delta_1 \psi, \delta_2 \psi) 
\equiv 
\delta_2 \Bigl( \varepsilon_{b c_1 \cdots c_{n-1}} \, 
\Theta^b(\psi; \delta_1 \psi) \Bigr) 
- \delta_1 \Bigl( \varepsilon_{b c_1 \cdots c_{n-1}} \, 
\Theta^b(\psi; \delta_2 \psi) \Bigr) ,  
\label{eqn:omegaDef} 
\end{equation} 
from the surface term $\Theta^b(\psi; \delta \psi)$ in the variation 
\begin{equation}
\delta \, \Bigl( \varepsilon_{c_1 \cdots c_{n}} \: L(\psi) 
\Bigr) = \varepsilon_{c_1 \cdots c_{n}} \, E_{I} \: \delta \psi^{I} 
+ \varepsilon_{c_1 \cdots c_{n}} \,  
\nabla_b \, \Theta^{b}(\psi; \delta \psi) ,   
\label{eqn:ActionVariation}
\end{equation} 
of the Lagrangian density $\varepsilon_{c_1 \cdots c_{n}} \: L(\psi)$,  
where $E_I = 0$ denotes the field equations. 

However, we should impose boundary conditions on $\psi^I$ 
so that the conserved charges $\chg[\psi, \zeta]$ are integrable. 
The integrability condition of $\chg[\psi, \zeta]$ is then written \cite{WaldZoupas} as  
\begin{equation} 
\int_{\partial {\cal C}} \zeta^{b} \, 
\omega_{b c_1 \cdots c_{n - 2}}(\psi; \delta_1 \psi, \delta_2 \psi) = 0 ,  
\label{eqn:IntegrabilityCond} 
\end{equation} 
for arbitrary independent variations $\delta_1 \psi^I$ and 
$\delta_2 \psi^I$ tangent to the covariant phase space. 
When the integrability condition (\ref{eqn:IntegrabilityCond}) is satisfied, 
the conserved charges $\chg[\psi, \zeta]$ 
are expressed in the form of a surface integral as 
\begin{equation} 
\chg[\psi; \zeta] = \int_{\partial {\cal C}} \frac{1}{2} \: 
\varepsilon_{b a c_1 \cdots c_{n-2}} 
\Bigl[ Q^{b a}(\psi; \zeta) + 2 \: \zeta^{[ b} \, B^{a ]}(\psi) \Bigr] 
+ \chg_{0}[\zeta] , 
\label{eqn:ChargeIntgrated} 
\end{equation} 
where $\chg_{0}[\zeta]$ is an integration constant, 
which we tentatively set to zero, 
$B^a(\psi)$ is defined by 
\begin{equation} 
\delta \int_{\partial {\cal C}} 
\varepsilon_{b a c_1 \cdots c_{n-2}} \, 
\zeta^{b} \, B^{a}(\psi) \equiv
\int_{\partial {\cal C}} \varepsilon_{b a c_1 \cdots c_{n-2}} \,  
\zeta^{b} \, \Theta^{a}(\psi; \delta \psi) , 
\label{eqn:BDef} 
\end{equation} 
and $Q^{b a}(\psi; \zeta)$ is given, by using $E_I = 0$, 
as the potential of the Noether current 
$J^b(\psi; \zeta) \equiv \Theta^b(\psi; {\cal L}_{\zeta} \psi) - \zeta^b L$, 
i.e., $J^b(\psi; \zeta) = \nabla_a Q^{b a}(\psi, \zeta)$.  

Under the integrability condition (\ref{eqn:IntegrabilityCond}), 
it has been also shown \cite{Koga01} that 
the Poisson brackets between $\chg[\psi, \zeta]$ are given by  
\begin{equation} 
\Bigl\{ \chg[\psi; \zeta_1] \; , \; \chg[\psi; \zeta_2] \Bigr\} = 
\chg[\psi; {\cal L}_{\zeta_1} \zeta_2] 
+ \ctrext[\zeta_1 , \zeta_2]  , 
\label{eqn:PissonBraGen} 
\end{equation} 
where the central term $\ctrext[\zeta_1 , \zeta_2]$ is evaluated as  
\begin{equation} 
\ctrext[\zeta_1 , \zeta_2] = \delta_{\zeta_2} \chg[\bfld; \zeta_1] 
- \chg[\bfld; {\cal L}_{\zeta_1} \zeta_2] ,  
\label{eqn:EvaluateK} 
\end{equation}  
on a background configuration $\bfld^I$, and 
$\delta_{\zeta_2} \chg[\bfld; \zeta_1]$ is calculated by   
\begin{eqnarray} 
\delta_{\zeta_2} \chg[\bfld; \zeta_1] 
& = & 
\int_{\partial {\cal C}} \frac{1}{2} \: \beps \Bigl[ {\cal L}_{\zeta_2} 
Q^{b a}(\bfld; \zeta_1) - {\cal L}_{\zeta_1} Q^{b a}(\bfld; \zeta_{2}) 
- Q^{b a}(\bfld; {\cal L}_{\zeta_2} \zeta_1) \nonumber \\ 
& & 
+ Q^{b a}(\bfld; \zeta_1) \, \bigl( \bder_{d} \, \zeta^d_2 \bigr) 
- Q^{b a}(\bfld; \zeta_2) \bigl( \bder_d \, \zeta^d_1 \bigr) 
+ 2 \: \zeta^{[ b}_1 \, \zeta^{a ]}_2 \, L(\bfld) \Bigr] . 
\label{eqn:CentralCharge} 
\end{eqnarray}  
The same expression for the central term has been derived by 
Silva \cite{Silva02}. Although Barnich and Brandt \cite{BarnichBrandt02} 
gave a slightly different expression, the difference is shown to vanish, 
when $\mathcal{L}_{\zeta} \bg_{a b}$ is vanishing on $\partial \mathcal{C}$, 
as we consider in this paper.  

In order to analyze the integrability condition (\ref{eqn:IntegrabilityCond}) 
and the central term (\ref{eqn:EvaluateK})  
for the actions of the asymptotic Killing vectors on the Killing horizon of 
the four-dimensional Kerr--Newman black hole, 
we now let $\zeta^a$ be given by (\ref{eqn:ZetaGenDefPre}) and $\bfld^I$ be 
the Kerr--Newman black hole solution. Correspondingly, we consider 
the four-dimensional Einstein--Maxwell theory, 
and thus the Lagrangian $L(\psi)$ is given by 
\begin{equation} 
L(\psi) = \frac{1}{16 \pi} \biggl[ R - F^{a b} F_{a b} \biggr] .  
\label{eqn:LagrangianEM}
\end{equation} 
As the dynamical field variables $\psi^I$, we take 
the metric $g_{a b}$ and the gauge potential $A_{a}$, 
because the variation of the Lagrangian (\ref{eqn:LagrangianEM}) 
with respect to these variables yields the field equations $E_I = 0$, 
from which we also find that $\Theta^b(\psi; \delta \psi)$ is given as  
\begin{equation} 
\Theta^b(\psi; \delta \psi) 
= \frac{1}{16 \pi} \biggl[ g_{a c} \nabla^b \delta g^{a c} - \nabla_a \delta g^{b a} 
- 4 F^{b a} \delta A_a \biggr] . 
\label{eqn:ThetaEM} 
\end{equation} 
The background configuration $\bfld^I$ is thus described by  
the metric $\bg_{a b}$, along with the regular gauge potential 
$\ba_a^{\nep}$ in the non-extreme case, and $\ba_a^{\eip}$ or $\ba_a^{\eiip}$ 
in the extreme case, as we presented in section \ref{sec:background}. 
While we focus in this paper on the spacetime exterior to the horizon of 
the Kerr--Newman black hole, 
the boundary $\partial {\cal C}$ of ${\cal C}$ is considered to consist of 
a horizon sphere only, 
by assuming that all the asymptotic Killing vectors fall off rapidly 
on the other boundary of ${\cal C}$ at infinity. 
In order to complete the construction of the covariant phase space, 
we now specify the behavior of the metric and the gauge potential 
near $\partial {\cal C}$  
so that the integrability condition (\ref{eqn:IntegrabilityCond}) is satisfied. 

%%%%%%%%%%%%%%%%%%%%%%%%%%%%%%%%%
% Boundary condition
%%%%%%%%%%%%%%%%%%%%%%%%%%%%%%%%%
\subsection{Boundary condition}

Since the symplectic current density 
$\omega_{b c d}(\psi; \delta_1 \psi, \delta_2 \psi)$ is bi-linear 
in $\delta \psi^I$ or their derivatives, 
the integrability condition (\ref{eqn:IntegrabilityCond}) is satisfied,  
if we impose the boundary condition that $\delta \psi^I$ and their derivatives 
fall off rapidly enough near the horizon sphere $\partial {\cal C}$. 
However, we are analyzing here the actions of the asymptotic symmetries  
in the covariant phase space, and thus we wish to impose a boundary condition 
so that any configurations of $\psi^I$ achieved 
by the asymptotic symmetry transformations 
reside within the covariant phase space. 

The desired boundary condition on the metric $g_{a b}$ is easily specified.   
We note that (\ref{eqn:AsymKilEq}) gives 
\begin{equation} 
\delta_{\zeta} g_{a b} = {\cal L}_{\zeta} g_{a b} = \odr(u) , 
\label{eqn:VarAKVMetric} 
\end{equation} 
as long as $g_{a b}$ takes the form of (\ref{eqn:PerturbMetric}).  
On the other hand, (\ref{eqn:PerturbMetric}) is described in the variational form as 
\begin{equation} 
\delta g_{a b} = \odr(u) . 
\label{eqn:BoundCondVarMetric} 
\end{equation} 
We note that (\ref{eqn:VarAKVMetric}) is consistent with 
(\ref{eqn:BoundCondVarMetric}), 
and hence the actions of all the asymptotic Killing vectors $\zeta^a$ 
are realized within a phase space, when the phase space is constructed under 
the boundary condition (\ref{eqn:BoundCondVarMetric}). 
In addition, it has been shown \cite{Koga01} that 
if (\ref{eqn:BoundCondVarMetric}) is imposed, 
the integrability condition (\ref{eqn:IntegrabilityCond}) 
of the conserved charges is satisfied in vacuum Einstein gravity. 
Also in presence of the electromagnetic field, we then naturally impose 
(\ref{eqn:BoundCondVarMetric}) as the boundary condition on the metric $g_{a b}$. 

In order to specify the boundary condition on the gauge potential $A_a$,  
we first need to incorporate $A_a$ 
into the notion of the asymptotic symmetries, so that the asymptotic Killing vectors 
$\zeta^a$ act on $A_a$ as the generators of the asymptotic symmetries. 
Since the background gauge potential $\ba_a$ is invariant along 
the Killing vectors $\KV^a$ on the horizon, 
as it is seen from (\ref{eqn:ConstPotNE}) and (\ref{eqn:ConstPotEx}), 
and the asymptotic Killing vectors are local generalization of 
Killing vectors, it might seem reasonable to require 
that $\ba_a$ is invariant also along the asymptotic Killing vectors $\zeta^a$ 
as ${\cal L}_{\zeta} \ba_a = \odr(u)$, and hence $\delta_{\zeta} \ba_a = \odr(u)$. 
By noting $\zeta^a \nabla_a u = \odr(u)$ and 
following the same argument as we used above for the boundary condition on the metric, 
it then would be natural to impose the boundary condition on the gauge potential 
as $\delta A_a = \odr(u)$. 
This would be plausible also from the viewpoint of integrability 
of the conserved charges $\chg[\psi, \zeta]$, since the integrability condition 
(\ref{eqn:IntegrabilityCond}) is satisfied when $\delta A_a = \odr(u)$, 
as we will see from (\ref{eqn:IntCondA}) below. 

However, the electromagnetic field has the gauge degree of freedom. 
We thus write as $\delta A_a = \nabla_a \chi + \odr(u)$, 
where $\chi$ may be arbitrary at this stage but will be 
constrained from the integrability condition in what follows. 
By using (\ref{eqn:ThetaEM}) and substituting (\ref{eqn:ZetaGenDefPre}) 
and (\ref{eqn:BoundCondVarMetric}) into (\ref{eqn:IntegrabilityCond}), 
we find that the integrability condition is described as 
\begin{equation} 
\int_{\partial {\cal C}} \varepsilon_{c d} \: 
\omega \biggl[ \bigl( \delta_1 F^{b a} \bigr) {\HKV}_b \bigl( \delta_2 A_a \bigr) 
- \bigl( \delta_2 F^{b a} \bigr) {\HKV}_b \bigl( \delta_1 A_a \bigr) \biggr] = 0 ,   
\label{eqn:IntCondA} 
\end{equation} 
where $\varepsilon_{c d}$ is the volume element on $\partial \mathcal{C}$ 
and we note that $\RKV^a$ are tangent to $\partial {\cal C}$. 
Since $\omega$ is an arbitrary function, it then follows that the inside of 
the bracket in (\ref{eqn:IntCondA}) should vanish 
in order that the integrability condition is 
satisfied for all the asymptotic Killing vectors.  
We see that it does vanish if we impose the condition $\chi = \odr(u)$ near the horizon, 
because $F^{a b}$ is antisymmetric and 
$\delta A_a = \nabla_{a} \chi + \odr(u) \propto \nabla_{a} u + \odr(u) 
\propto {\HKV}_a + \odr(u)$ in that case. 
Imposing a boundary condition on $\chi$ might look strange, 
but it is actually natural that a boundary condition on the metric is 
combined with a condition on gauge transformations on the boundary, as it occurs 
in the case of an asymptotically anti-de Sitter spacetime \cite{HenneauxTeitelboim85}, 
for example. 
Therefore, we impose the boundary condition on $A_a$ as 
\begin{equation} 
\delta A_a = \nabla_a ( u \, \chi' ) + \odr(u) , 
\label{eqn:BoundCondVarA} 
\end{equation} 
and hence $A_a$ is written as 
\begin{equation} 
A_a = \ba_a + \nabla_a ( u \, \chi' ) + \odr(u) , 
\label{eqn:BoundCondA} 
\end{equation} 
where $\chi'$ is an arbitrary smooth function. 

One can see that the boundary condition (\ref{eqn:BoundCondVarA}) is supported 
by the natural requirement that the conserved charges $\chg[\psi; \zeta]$ 
should not depend on the electromagnetic gauge. 
Suppose that one of the variations of $\psi^I$ 
in the arguments of the symplectic current density 
$\omega_{c d e}(\psi; \delta_1 \psi , \delta_2 \psi)$, 
say $\delta_2 \psi^I$, is a pure gauge transformation as 
$\delta_2 g_{a b} = 0$ and $\delta_2 A_a = \nabla_a \chi$,  
and the other is given by the Lie derivatives along an asymptotic Killing vector 
$\zeta^a$ as 
$\delta_1 g_{a b} = {\cal L}_{\zeta} g_{a b}$ and $\delta_1 A_a = {\cal L}_{\zeta} A_a$.  
Then, the variation $\delta_2 \chg[\psi; \zeta]$ of $\chg[\psi; \zeta]$ 
under the gauge transformation $\delta_2$ is calculated, 
by using (\ref{eqn:ChargeDef}) and the Maxwell equation, as 
\begin{equation} 
\delta_2 \chg[\psi; \zeta] =
\frac{1}{8 \pi} \int_{\partial {\cal C}} 
\epsilon_{b a c d} \: \chi 
\Bigl[ {\cal L}_{\zeta} F^{b a} 
+ \frac{1}{2} g^{d e} 
( {\cal L}_{\zeta} g_{d e} ) F^{b a} \Bigr] . 
\end{equation} 
Although ${\cal L}_{\zeta} g_{a b} = \delta_{\zeta} g_{a b}$ vanishes 
on the horizon sphere $\partial {\cal C}$, 
due to the boundary condition (\ref{eqn:BoundCondVarMetric}), 
${\cal L}_{\zeta} F^{a b}$ does not, in general. 
If we impose $\chi = \odr(u)$, however, we have 
$\delta_2 \chg[\psi; \zeta] = 0$, as required. 
Thus, we find, also from gauge invariance of the conserved charges, 
that the boundary condition (\ref{eqn:BoundCondVarA}) is reasonable. 

While the boundary condition (\ref{eqn:BoundCondVarA}) on $A_a$ 
has been determined by incorporating the electromagnetic field into 
the asymptotic symmetries and imposing the integrability condition 
of the conserved charges, 
it does not necessarily ensure that this boundary condition is preserved 
under the actions of the asymptotic Killing vectors. 
When we take the Lie derivatives of $A_a$ along 
the asymptotic Killing vectors $\zeta^a$, we indeed find 
\begin{equation} \fl 
\delta_{\zeta} A_a = {\cal L}_{\zeta} A_a = 
\HKV^b A_b \nabla_a \omega + \nabla_a \Bigl[ \omega \HKV^b \nabla_b ( u \, \chi' ) 
+ a^{\scriptscriptstyle (i)} \RKV^b \nabla_b ( u \, \chi' )  
- \potkil A_b \nabla^b \omega \Bigr] + \odr(u) , 
\label{eqn:GenATrans}  
\end{equation}  
from (\ref{eqn:ZetaGenDefPre}) and (\ref{eqn:BoundCondA}). 
We see that the second term in (\ref{eqn:GenATrans}) is the gradient of 
a scalar of $\odr(u)$, because we have 
$\HKV^b \nabla_b u = \odr(u)$ and $\RKV^b \nabla_b u = \odr(u)$. 
Then, by noting that (\ref{eqn:BoundCondVarA}) yields $n^a \delta A_a = \odr(u)$ 
for an arbitrary vector $n^a$ tangent to $u = 0$, 
a necessary condition for $\delta_{\zeta} A_a$ 
to take the form of (\ref{eqn:BoundCondVarA}) 
is found as  
\begin{equation} 
\HKV^b A_b \: n^a \nabla_a \omega = \odr(u) .  
\end{equation} 
In order that any configurations of $A_a$ 
achieved by the asymptotic symmetry transformations reside in the covariant 
phase space, we thus impose either  
\begin{equation} 
n^a \nabla_a \omega = \odr(u) , 
\label{eqn:CondRedGroup} 
\end{equation} 
or 
\begin{equation} 
\HKV^b A_b = \odr(u) . 
\label{eqn:ClosureCondA} 
\end{equation} 
When (\ref{eqn:ClosureCondA}) is satisfied, (\ref{eqn:GenATrans}) is consistent 
with the boundary condition (\ref{eqn:BoundCondVarA}) for arbitrary $\omega$, 
and hence the actions of all the asymptotic Killing vectors are 
appropriately represented in the covariant phase space in this case.  
If (\ref{eqn:CondRedGroup}) is satisfied, however, 
$\omega$ must be constant on $u = 0$. 
Moreover, as we have seen from (\ref{eqn:UDependenceOm}), 
dependence of $\omega$ on $u$ cancels out up to $\odr(u^2)$, 
which occurs also in (\ref{eqn:GenATrans}). 
Thus, $\omega$ in this case must be given as $\omega = const. + \odr(u^2)$, 
and then the asymptotic symmetry group is nothing more than 
the isometry group. 
On the other hand, from (\ref{eqn:GenATrans}), we have $\delta_{\zeta} A_a = \odr(u)$  when $\omega = const. + \odr(u^2)$, which indicates that 
the isometry group is represented appropriately in the covariant phase space 
under the boundary condition (\ref{eqn:BoundCondVarA}). 
Therefore, we see that the asymptotic symmetry group is 
necessarily reduced to the isometry group in the covariant phase space, 
when the condition (\ref{eqn:ClosureCondA}) is not satisfied and hence 
(\ref{eqn:CondRedGroup}) must be imposed. 

%%%%%%%%%%%%%%%%%%%%%%%%%%%%
\subsection{Degeneracy} 

In the case of the non-extreme Kerr--Newman black hole, 
the regular gauge potential $\ba_a^{\nep}$ satisfies (\ref{eqn:GaugeCondNE}), 
and hence (\ref{eqn:ClosureCondA}) follows from (\ref{eqn:BoundCondA}). 
Indeed, whenever the surface gravity is non-vanishing, 
(\ref{eqn:ClosureCondA}) is shown to be satisfied on an arbitrary Killing horizon 
in a regular gauge that satisfies the boundary condition (\ref{eqn:BoundCondVarA}). 
A Killing horizon with the non-vanishing surface gravity has 
a bifurcation surface \cite{RatzWald}, 
where the Killing vector $\HKV^a$ vanishes, and so does $\HKV^b A_b$. 
In addition, since the boundary condition (\ref{eqn:BoundCondVarA}) implies 
\begin{equation} 
{\cal L}_{\HKV} A_a = \delta_{\HKV} A_a = \nabla_a ( u \, \chi' ) + \odr(u) ,   
\end{equation} 
we have 
\begin{equation} 
{\cal L}_{\HKV} \HKV^a A_a = \odr(u) .  
\end{equation} 
Therefore, $\HKV^a A_a$ vanishes all over the horizon $u = 0$, 
and (\ref{eqn:ClosureCondA}) holds generally on a Killing horizon 
with the non-vanishing surface gravity.  
The actions of all the asymptotic Killing vectors are thus represented 
in the covariant phase space, without reduction of the asymptotic symmetry group. 

However, an extreme black hole does not possess a bifurcation surface, 
and hence the above argument in the case of non-vanishing surface gravity 
does not apply to an extreme black hole. 
It is possible even in the extreme case to choose a gauge 
which imitates that of a non-extreme black hole, in the sense that 
(\ref{eqn:ClosureCondA}) is satisfied, such as $\ba_a^{\eip}$. 
However, there are no reasons to exclude the gauges 
that do not satisfy (\ref{eqn:ClosureCondA}), such as $\ba_a^{\eiip}$,   
because the Killing vector $\HKV^a$ does not 
vanish anywhere on the horizon, as we see from (\ref{eqn:XiHexpEX}). 
In the latter case, we need to impose (\ref{eqn:CondRedGroup}), and hence 
the asymptotic symmetry group is necessarily reduced to the isometry group. 
We emphasize again that the gauges that do not satisfy (\ref{eqn:ClosureCondA}) 
reflect faithfully the global geometric structure of the Killing horizon of  
an extreme black hole, i.e., absence of a bifurcation surface.  
In other words, the asymptotic symmetry group must be reduced 
to the isometry group when we respect this global geometric structure 
particular to an extreme black hole. 

Here we recall that an asymptotic Killing horizon \cite{Koga-a} 
was considered as a local geometric structure generalized from the notion of 
a Killing horizon, and is defined by the pair $( H , \HGV^a )$ 
of a null hypersurface $H$ and its generator $\HGV^a$. 
The generator $\HGV^a$ of an asymptotic Killing horizon 
is then given by the asymptotic Killing vectors that 
become null on $H$, i.e., those with $a^{\scriptscriptstyle (i)} = 0$ 
in (\ref{eqn:ZetaGenDefPre}). 
Since the function $\omega$ in (\ref{eqn:ZetaGenDefPre}) is arbitrary, 
there exist infinitely many asymptotic Killing horizons 
on the common null hypersurface $H$. 
In particular, an arbitrary Killing horizon is accompanied by infinitely many 
asymptotic Killing horizons, and hence this provides 
degeneracy associated with the local structure of a Killing horizon. 
However, when the asymptotic symmetry group is reduced 
to the isometry group, which occurs in the case of an extreme black hole, 
the degeneracy of the asymptotic Killing horizons disappears, 
leaving the extreme black hole isolated. 

It was also argued in \cite{Koga-a}, 
based on the behavior of the acceleration associated with asymptotic Killing horizons, 
that the Killing horizon of an extreme black hole shows 
two contrastive aspects. 
On one hand, the Killing horizon of an extreme black hole is considered as isolated 
from the asymptotic Killing horizons on itself, in contrast with the Killing horizon 
of a non-extreme black hole, which is continuously deformed into 
asymptotic Killing horizons.  
On the other hand, infinitely many asymptotic Killing horizons reside 
on the Killing horizon both of a non-extreme black hole and of an extreme black hole, 
and hence an extreme black hole carries as many 
asymptotic Killing horizons as a non-extreme (near-extreme) black hole.  
The former aspect is manifest when the global structure near the horizon is probed, 
while the latter is relevant to the local structure. 
Therefore, an extreme black hole looks isolated 
when the global structure of its Killing horizon is respected, but it behaves similarly to 
a non-extreme black hole as far as its local structure is concerned. 
We also recall that the Euclidean approach to black hole thermodynamics 
shows that the entropy of an extreme black hole vanishes 
\cite{HawkingHR95,Teitelboim95}, 
which implies that there exists only a single microscopic state, 
while it is shown in string theory to obey the Bekenstein--Hawking formula 
\cite{StromingerVafa96,Horowitz96,Sen05}, indicating 
that an extreme black hole is accompanied by microscopic states 
as many as a non-extreme black hole. 
We now see that the behavior of the asymptotic Killing horizons \cite{Koga-a}, 
which was argued based on the acceleration and found consistent with 
both of these reliable results in string theory and the Euclidean approach, 
is realized also in the covariant phase space. 
The covariant phase space of an extreme black hole 
constructed under the boundary condition that imitates a non-extreme black hole 
has the size comparable with that of a non-extreme black hole, 
but the boundary condition that reflects faithfully the global geometric 
structure of an extreme black hole yields the covariant phase space 
where degeneracy of asymptotic Killing horizons is not allowed. 

We now turn to the issue of the algebra defined by 
the Poisson brackets (\ref{eqn:PissonBraGen}). 
In particular, it is meaningful to see whether a non-vanishing central charge arises 
in the Poisson brackets algebra, i.e., whether the sub-group of 
diffeomorphisms described by the asymptotic Killing vectors 
is represented with an anomaly. 
Since we have imposed the boundary condition 
so that the conserved charges $\chg[\psi, \zeta]$ conjugate to 
the asymptotic Killing vectors $\zeta^a$ are integrable, 
we can compute the central term $\ctrext[\zeta_1 , \zeta_2]$ 
by (\ref{eqn:EvaluateK}) and (\ref{eqn:CentralCharge}), 
where $Q^{b a}(\psi;\zeta)$ is found from the Lagrangian (\ref{eqn:LagrangianEM}) as 
\begin{equation} 
Q^{b a}(\psi;\zeta) = \frac{1}{16 \pi} \biggl[ \nabla^a \zeta^b 
- \nabla^b \zeta^a + 4 F^{a b} A_d \zeta^d \biggr] . 
\label{eqn:QExplicit} 
\end{equation} 
We then substitute (\ref{eqn:ZetaGenDefPre}) and (\ref{eqn:QExplicit}) 
into (\ref{eqn:CentralCharge}) and use the field equations. 
We also note the facts that 
$\HKV^a$ and $\RKV^a$ are Killing vectors, 
which satisfy (\ref{eqn:FSymmetry}) and commute with each other, 
that the horizon sphere $\partial {\cal C}$ does not have its own boundaries, 
and that $\HKV^a$ is hypersurface orthogonal.  
Then, under the boundary condition imposed above, 
whether the asymptotic symmetry group 
is reduced ($\omega = const. + \odr(u^2)$) or not ($\HKV^a A_a = \odr(u)$), 
a lengthy calculation yields 
\begin{equation} 
\delta_{\zeta_2} \chg[\bfld; \zeta_1] = 0 .  
\label{eqn:VanishingC} 
\end{equation}   
Therefore, the Poisson brackets algebra of the conserved charges $\chg[\psi, \zeta]$ 
is given by 
\begin{equation} 
\Bigl\{ \chg[\psi; \zeta_1] \; , \; \chg[\psi; \zeta_2] \Bigr\} = 
\chg[\psi; {\cal L}_{\zeta_1} \zeta_2] - \chg[\bfld; {\cal L}_{\zeta_1} \zeta_2] .  
\label{eqn:FaulseCentral} 
\end{equation} 
When we redefine $\chg[\psi; \zeta]$ by an additive constant as 
\begin{equation} 
\chg[\psi; \zeta] \rightarrow 
\chg'[\psi; \zeta] \equiv \chg[\psi; \zeta]- \chg[\bfld; \zeta] , 
\end{equation} 
which is achieved by adjusting the integration constant $\chg_{0}[\zeta]$ 
in (\ref{eqn:ChargeIntgrated}), 
we see that (\ref{eqn:FaulseCentral}) is rewritten as 
\begin{equation} 
\Bigl\{ \chg'[\psi; \zeta_1] \; , \; \chg'[\psi; \zeta_2] \Bigr\} 
= \chg'[\psi; {\cal L}_{\zeta_1} \zeta_2] . 
\label{eqn:TrueCentral} 
\end{equation} 
This shows that the central charge vanishes, as in the case of \cite{Koga01}, 
and hence that there arises no anomaly of diffeomorphism invariance.  
In particular, while the Lie brackets algebra of the asymptotic Killing vectors $\zeta^a$ 
contains the $\mathit{diff}(S^1)$ or $\mathit{diff}(R^1)$ sub-algebra \cite{Koga-a}, 
the corresponding Poisson brackets sub-algebra does not possess a non-vanishing 
central charge. 

We note that the vanishing central charge in the Poisson brackets algebra 
is an immediate consequence of (\ref{eqn:VanishingC}). However, we can 
argue that (\ref{eqn:VanishingC}) serves also as an evidence 
that the microscopic states of black hole thermodynamics are 
described by the asymptotic Killing horizons. To see this, we set as 
\begin{equation} 
\zeta_1^a = \HKV^a , \qquad 
\zeta_2^a = \HGV^a \equiv \omega \: \HKV^a - \potkil \nabla^a \omega + \odr(u) , 
\label{eqn:SetVecs} 
\end{equation}  
and substitute into (\ref{eqn:VanishingC}). We then have   
\begin{equation} 
\delta_{\HGV} \chg[\bfld; \HKV] = 0 . 
\label{eqn:InvNoether} 
\end{equation} 
Since the variation $\delta$ is defined to act only on the dynamical field variables 
$\psi^I$, as we see from the variation of the Lagrangian (\ref{eqn:ActionVariation}), 
but not on $\HKV^a$, 
(\ref{eqn:InvNoether}) indicates that the conserved charge $\chg[\bfld; \HKV]$ 
remains unaffected under the infinitesimal transformations 
\begin{equation} 
\psi^I \rightarrow \psi^I + {\cal L}_{\HGV} \psi^I , \qquad 
\HKV^a \rightarrow \HKV^a ,  
\label{eqn:DynTrans} 
\end{equation} 
i.e., the transformations where the dynamical field variables $\psi^I$ 
are transformed by the Lie derivatives but the vector $\HKV^a$ is fixed. 
However, we see that the transformations (\ref{eqn:DynTrans}) are equivalent, 
up to diffeomorphisms, 
to the transformations where $\psi^I$ are fixed but $\HKV^a$ is transformed as 
\begin{equation} 
\psi^I \rightarrow \psi^I , \qquad 
\HKV^a \rightarrow \HKV^a - {\cal L}_{\HGV} \HKV^a , 
\label{eqn:HorDef} 
\end{equation} 
because the transformations (\ref{eqn:DynTrans}) 
followed by the infinitesimal diffeomorphisms 
\begin{equation} 
\psi^I \rightarrow \psi^I - {\cal L}_{\HGV} \psi^I , \qquad 
\HKV^a \rightarrow \HKV^a - {\cal L}_{\HGV} \HKV^a , 
\end{equation} 
reduce to (\ref{eqn:HorDef}). 
On the other hand, $\chg[\bfld; \HKV]$ (with $\chg_{0}[\HKV] = 0$) 
has been shown \cite{FirstLaw} 
to coincide with $T S$ on a stationary and axisymmetric black hole spacetime, 
where $T$ and $S$ denote the temperature and the entropy of the black hole, 
respectively. 
The entropy of the Kerr--Newman black hole is given by the Bekenstein--Hawking 
formula, and the Lie derivative of the volume element of a horizon sphere vanishes 
along the null direction on the horizon. 
Since the zeroth law holds and $\HGV^a$ is null on the horizon, 
we then find that $\chg[\bfld; \HKV] = T S$ is 
invariant under the diffeomorphisms along $\HGV^a$, and hence 
that it is left unchanged also under the transformations (\ref{eqn:HorDef}). 
Furthermore, we can write \cite{Koga-a} as  
\begin{equation} 
\HKV^a - {\cal L}_{\HGV} \HKV^a 
= \omega' \HKV^a - \potkil \nabla^a \omega' + \odr(u) , 
\end{equation} 
where $\omega' = 1 + \HKV^a \nabla_a \omega$, 
and thus $\HKV^a - {\cal L}_{\HGV} \HKV^a$ is found to be 
an asymptotic Killing vector that becomes null on the horizon. As we mentioned above, 
such an asymptotic Killing vector generates an asymptotic Killing horizon 
on the same null hypersurface as the Killing horizon.  
Therefore, the Killing horizon is transformed 
into asymptotic Killing horizons under the transformations (\ref{eqn:HorDef}), 
and thus we see that the value of the conserved charge $\chg[\bfld; \HKV]$ is left 
unchanged under these transformations from the Killing horizon into 
the asymptotic Killing horizons. 
Since $\chg[\bfld; \HKV]$ gives the thermodynamical quantity 
and the function $\omega$ in (\ref{eqn:SetVecs}) is arbitrary,  
it then implies that all the asymptotic Killing horizons 
infinitesimally transformed 
from the Killing horizon are associated with the same 
macroscopic (thermodynamical) state. 
This may suggests that the asymptotic Killing horizons are 
regarded as degenerate from a macroscopic point of view, 
while they are physically distinguishable \cite{Koga-a}, 
as we mentioned in Introduction.  

%%%%%%%%%%%%%%%%%%%%%%%%%%%%%%%%% 
%  Summary and discussion
%%%%%%%%%%%%%%%%%%%%%%%%%%%%%%%%%
\section{Summary and discussion} 
\label{sec:discussion} 

We first derived the general form of the asymptotic Killing vectors on 
the Killing horizon of the Kerr--Newman black hole, 
which was actually shown to be possessed universally 
by arbitrary Killing horizons \cite{Koga-a}. 
One might suspect that the asymptotic symmetries generated by these 
asymptotic Killing vectors are nothing more than 
a sort of gauge, but it should be emphasized that asymptotic Killing horizons, 
which are described by a sub-group of these asymptotic symmetries, 
are distinguished physically by the acceleration associated with them \cite{Koga-a}. 
It is then natural to expect that asymptotic Killing horizons 
have something to do with the universal physics of a Killing horizon, 
in particular, its thermal feature. 

We then considered the covariant phase space associated with 
these asymptotic Killing vectors. 
By incorporating the electromagnetic field 
into the asymptotic symmetries, we found that 
two types of the boundary condition on the gauge potential are possible 
in the case of an extreme black hole, 
and that they give the covariant phase spaces of an extreme black hole  
with different sizes. 
The boundary condition that imitates a non-extreme black hole provides  
the covariant phase space of the size comparable with a non-extreme black hole, 
while the covariant phase space resulting from 
the boundary condition that respects the global structure of a Killing horizon 
particular to an extreme black hole is small enough 
to exclude the degeneracy of asymptotic Killing horizons. 
It should be stressed, however, that what plays an essential role is 
not the electromagnetic gauge, but the geometric structure of 
the Killing horizon of an extreme black hole. 
The electromagnetic field simply probes this geometric structure. 
Even without the electromagnetic field, as in the case of 
the Kerr black hole, it is actually possible 
to reduce the covariant phase space of the asymptotic Killing vectors  
by imposing stringent conditions that respect the global geometric structure of 
a Killing horizon particular to an extreme black hole. 
Therefore, the behavior of the asymptotic Killing horizons 
of an extreme black hole, which was found from the acceleration associated with 
them \cite{Koga-a}, is properly realized also in the covariant phase space. 
In particular, the covariant phase space 
of the asymptotic Killing vectors is consistent with both 
the entropy of an extreme black hole in string theory 
\cite{StromingerVafa96,Horowitz96,Sen05} 
and that in the Euclidean approach \cite{HawkingHR95,Teitelboim95}, 
which are apparently conflicting with each other but might be viewed 
as dual aspects of a single phenomenon. 

It is important also to mention 
the issue of the central charge in the Poisson brackets algebra, 
which represents an anomaly of diffeomorphism invariance. 
We showed in this paper that the central charge vanishes. 
Thus, straightforward application of the same method as that 
applied to the \textit{infinity} of the B.T.Z. black hole spacetime \cite{Strominger98} 
does not reproduce the Bekenstein--Hawking formula. 
One might then consider that this result indicates that the asymptotic symmetries 
analyzed in this paper have nothing to do with microscopic states responsible 
for the thermal feature of a horizon, particularly if we persist in following 
the derivation in the case of the B.T.Z. black hole \cite{Strominger98}. 
However, there actually exists an explicit \textit{microscopic} model 
in the theory of induced gravity \cite{FrolovFZ03}, 
where the Bekenstein--Hawking formula has been reproduced 
based on the asymptotic symmetries 
with the \textit{vanishing total central charge}, 
while each microscopic field in this model carries a non-vanishing central charge. 
Moreover, in the recent approach to Hawking radiation \cite{RobinsonWilczek05}, 
it was found \cite{IsoUW} that the anomaly cancellation condition, 
which thus requires that 
\textit{anomalies of diffeomorphism and gauge invariance are absent}, 
along with appropriate conditions on the effective energy-momentum, 
lead to the Hawking flux with the correct temperature. 
Therefore, the vanishing central charge 
looks consistent with the thermal feature of a horizon, at least 
in an effective theory. 
Furthermore, we saw that the vanishing central charge provides an evidence 
that all the asymptotic Killing horizons on a Killing horizon are associated with 
the same macroscopic state, which may imply that 
the microscopic states of black hole thermodynamics are 
described by the asymptotic Killing horizons. 
Probably, not all of these asymptotic Killing horizon will contribute 
to black hole entropy, and a sort of quantization condition may be imposed 
in a quantum theory, 
which will allow only a discrete subset of the asymptotic Killing horizons 
to constitute a thermal object. 
It will be interesting to analyze this issue further in future investigations. 

%%%%%%%%%%%%%%%%%%%%%%%%%%%%%%%%%
% Acknowledgments 
%%%%%%%%%%%%%%%%%%%%%%%%%%%%%%%%%
\ack 
I would like to thank A. Hosoya, K. Maeda, M. Natsuume, 
G. Kang and M. Park for useful discussions and encouragement. 

%%%%%%%%%%%%%%%%%%%%%%%%%%%%%%%%%
%%%%%%%%%%%%%%%%%%%%%
%           Refs.
%%%%%%%%%%%%%%%%%%%%% 

%%%%%%%%%%%%%%%%%%%%

\end{document}